\documentclass[aps,prc,twocolumn,showpacs,superscriptaddress]{revtex4-2}
\usepackage{natbib}
\usepackage{graphicx}
\usepackage{color}
\usepackage{physics}
\usepackage{amsfonts}
\begin{document}
\bibliographystyle{revtex}
\title{Comparative analysis of formalisms and performances of three different beyond mean-field approaches}
\vspace{0.5cm}
\author{F. Knapp}   
\affiliation{Institute of Particle and Nuclear Physics, Faculty of Mathematics and Physics,  Charles University, V Hole\v sovi\v ck\'ach 2, 180 00 Prague, Czech Republic } 
\author{P. Papakonstantinou}
\affiliation{Rare Isotope Science Project, Institute for Basic Science, Daejeon 34000, Korea}
\author{P. Vesel\'y} 
\affiliation{Nuclear Physics Institute,
Czech Academy of Sciences, 250 68 \v Re\v z, Czech Republic} 
\author{G. De Gregorio}
\affiliation{Dipartimento di Matematica e Fisica, Universit$\grave{a}$ degli Studi della Campania "Luigi Vanvitelli",
viale Abramo Lincoln 5, I-81100 Caserta, Italy}
\affiliation{Istituto Nazionale di Fisica Nucleare, Complesso Universitario di Monte S. Angelo, Via Cintia, I-80126 Napoli, Italy}
\author{J. Herko}
\affiliation{Department of Physics and Astronomy, University of Notre Dame, Notre Dame, Indiana 46556-5670, USA}
\author{N. Lo Iudice}
\affiliation{Dipartimento di Fisica, 
Universit$\grave{a}$ di Napoli Federico II, 80126 Napoli, Italy} 
\date{\today}

\begin{abstract}
\noindent 
We investigate the differences and analogies between the equation of motion phonon method (EMPM) and  second Tamm-Dancoff and random-phase approximations (STDA and SRPA)
paying special attention to the problem of spurious center-of-mass (c.m.) admixtures. 
In order to compare them on an equal footing, we perform self-consistent calculations of the multipole strength distributions in selected doubly magic nuclei within a space including up to two-particle-two-hole (2p-2h) basis states using the UCOM two-body intrinsic Hamiltonian and we explore the tools each approach supplies for removing the spurious c.m. admixtures.
We find that the EMPM and STDA yield exactly the same results when the same intrinsic Hamiltonian is used and the coupling of the Hartree-Fock state with the 2p-2h space is neglected, 
but, unlike STDA and SRPA, the EMPM offers the possibility to completely remove c.m. admixtures. 
\end{abstract}
\maketitle

\section{Introduction}  
The random phase approximation (RPA)  has become the canonical approach to study the collective response of nuclei to external probes~\cite{RS80,Hav2001}.
It  provides a satisfactory description of the gross features of collective modes, such as their centroid energy and total strength, and
as a linear-response theory it is generally preferred over the simpler Tamm-Dancoff approximation (TDA). 
However, in order to describe more detailed properties, like fragmentation and damping of giant resonances (GR) beyond the Landau mechanism, it is necessary to go beyond the harmonic approximation underlying the RPA method  
and couple the particle-hole  (p-h) states building up the RPA  phonons to more complex configurations.  

The earliest extensions  were achieved within the  particle-vibration coupling (PVC) \cite{BoMo} and  quasiparticle-phonon models (QPM) \cite{Solov}.
The QPM adopts a separable interaction to generate quasiparticle RPA (QRPA) phonons and couples them to two and, in some cases, three RPA phonon
 configurations to describe low- and high-energy collective modes \cite{LoIudice2012}.    
 
In its first formulation \cite{BertchBortBr}, the PVC had a phenomenological character and was focused mainly on the fragmentation and damping of GR. It was then linked to energy density functional (EDF) theories and reformulated microscopically through the use of Skyrme interactions  \cite{NiuColo2014}. The same connection 
was established within the Green function (GF) framework  based on the time blocking approximation (TBA) where the PVC emerges from non-relativistic \cite{LitvTsel2007}  and relativistic \cite{LitvRingTse2007,LitvSchuck19,LitvZhang21} EDF.
Density dependent effective interactions derived from non-relativistic EDF, like Skyrme and Gogny, were also used in second RPA (SRPA) \cite{Gamba2015,GambaCo2012}.

 The SRPA, which also has a long history~\cite{Wam1988}, is the most straightforward  RPA extension. It solves directly the eigenvalue problem in a space spanned by p-h plus 2p-2h basis states.  An example is provided by the calculations \cite{PapaRot2009,PapaRot2010} performed by using  a potential obtained through the unitary correlation operator method (UCOM) \cite{Feldmeier1998}. 
 An even simpler approach is the second Tamm-Dancoff approximation (STDA) which is obtained from SRPA if the correlated ground state is replaced by 
the Hartree-Fock (HF) vacuum from the beginning, which in practice means that only the forward p-h and 2p-2h amplitudes are retained. 

The RPA extensions based on the phenomenological EDF theories have to deal with a double counting problem. By going beyond the mean-field approximation, correlations already present in the ground state may be induced since the parameters of the EDF are determined so as to reproduce the ground state properties within HF. In order to avoid such a redundancy it is necessary either to redetermine the parameters or to  adopt the so-called subtraction method proposed by Tselayev within the GFTBA \cite{Tsel2007,Tsel2013} and used also within the SRPA  \cite{Gamba2015}.

\begin{figure*}[ht!]
\includegraphics[width=\textwidth]{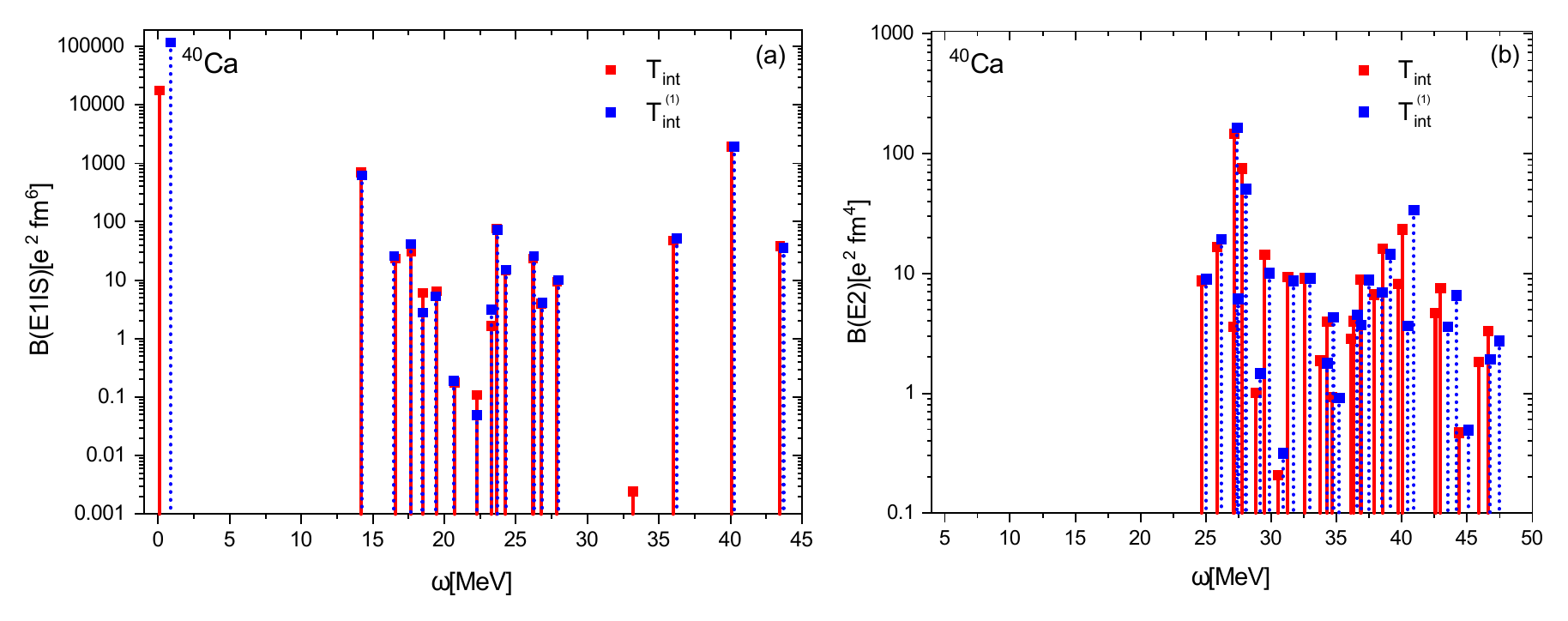}
\caption{(Color online) \label{fig:spurious} $E1$ (a) and $E2$ (b) responses of $^{40}$Ca evaluated in RPA with the intrinsic two-body kinetic energy operator $T_{int}$ ("intrinsic") and with the effective one-body kinetic energy operator $T_{int}^{(1)}$ ("effective").}
\end{figure*}

Another direct approach to the nuclear response is provided by  the equation of motion phonon method (EMPM)  \cite{AndLo,AndLo1}. 
 In  its upgraded formulation \cite{Bianco}, the TDA phonons are the basic constituents of an orthonormal basis of $n$-phonon $(n=2,3,4,....)$ states generated from solving iteratively a set of equations  of motion in each $n$-phonon subspace. These states together with the HF state $(n=0)$ and the TDA phonons $(n=1)$ are adopted to solve the full eigenvalue problem. The method was also formulated in the quasiparticle language suitable for open shell nuclei \cite{DeGreg16}  and  in the p(h)-phonon scheme for the study of odd-nuclei \cite{DeGreg16a,DeGreg17a,DeGreg17b,DeGreg18,DeGreg19,DeGregorio2020}.

\begin{figure*}[ht!]
\includegraphics[width=\textwidth]{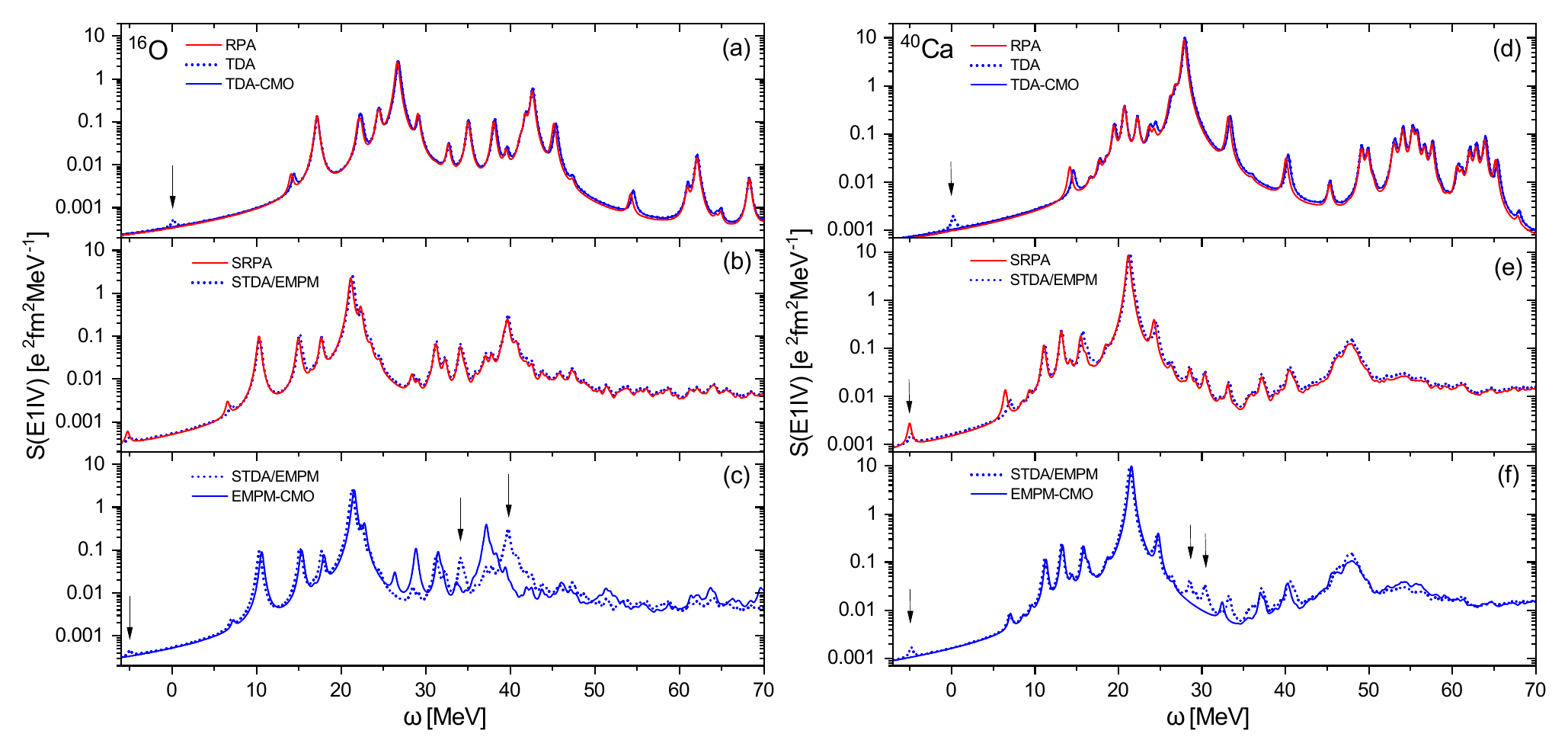}
\caption{(Color online) \label{fig1} Isovector $E1$ strength functions   in $^{16}$O and $^{40}$Ca  in different approaches.
The TDA strength is computed before (TDA) and after the application of Gram-Schmidt c.m. orthogonalization procedure (TDA-CMO). In this and the following figures, a single line is drawn for  STDA and EMPM since they yield identical results. The arrows indicate states with significant spurious components which disappear if c.m. SVD orthogonalization procedure (EMPM-CMO) is applied. 
}
\end{figure*}

\begin{figure*}[ht!]
\includegraphics[width=\textwidth]{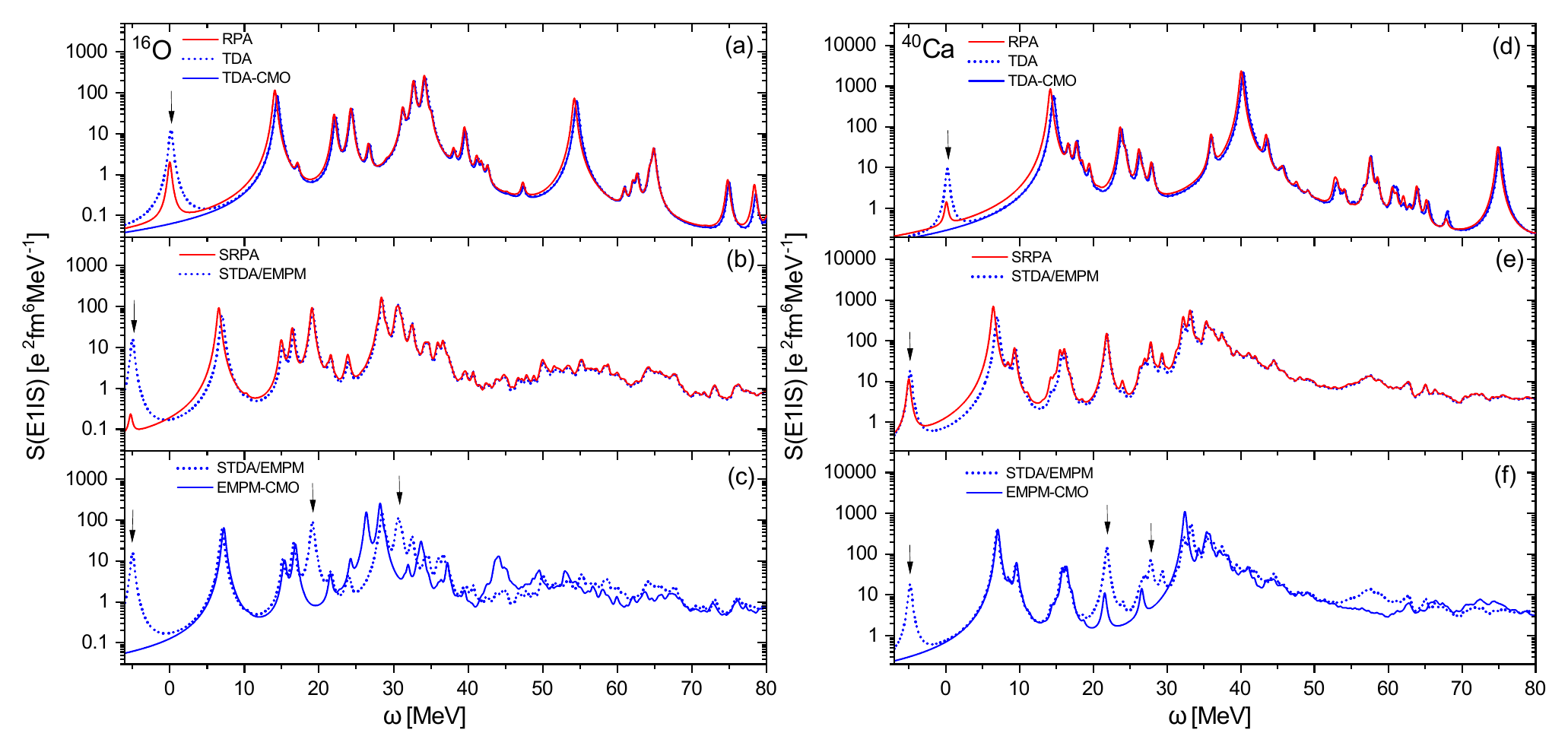}
\caption{(Color online) \label{fig2} Isoscalar $E1$ strength functions  in $^{16}$O and $^{40}$Ca.}
\end{figure*}

The EMPM does not rely on any approximation except for the truncation of the configuration space and the number of phonons. Therefore, we should expect that, within a space including 
 up to the two-phonon subspace, it should yield exactly the same results one obtains in STDA. This check represents the preliminary goal of the present work. In order to make a consistent comparison between the three different approaches, we neglect the coupling between HF and two-phonon states considered in the EMPM but neglected in STDA and SRPA. 
In all cases, the excitations are built on the HF ground state, which in turn is obtained with the same Hamiltonian as used for the couplings. In this sense, we call our calculations ``self-consistent." Furthermore, we take into account the interaction among the different 2p-2h configurations present in the EMPM but often neglected in both STDA and SRPA.  
 
 The scope of our study is wider. We intend to put on display analogies and differences between the three formalisms and to investigate if and how their specific features impact on their performances. 
 To this purpose, we will adopt the same UCOM two-body Hamiltonian throughout this work to determine the multipole response in some selected doubly-magic nuclei.

 We will pay special attention to the  problem  of the center of mass (c.m.) which may become critical once we go beyond the mean-field approximation.
 In fact,  we know that the decoupling between intrinsic and c.m. motion is achieved in RPA if a HF basis in a complete or large enough p-h space is adopted~\cite{Rowe1968}, while in SRPA this is generally not true in spite of Thouless's theorem~\cite{Rowe1968,Yannou87}, because the stability condition is violated~\cite{Papa14a}. 
In TDA, the decoupling is obtained   by exploiting the Gram-Schmidt orthogonalization method \cite{Bianco14}.
 In the  EMPM  the c.m. spurious admixtures can be removed from the whole multiphonon basis under no constraint  and for any single-particle (s.p.) basis  by a method which  exploits the singular value decomposition (SVD) \cite{DEGREGORIO2021,DeGreg2022}.    
 The comparison between the three approaches will enable us to establish the role of the c.m. motion on the different multipole responses and how important is the removal of such a motion.

\begin{figure*}[ht!]
\includegraphics[width=\textwidth]{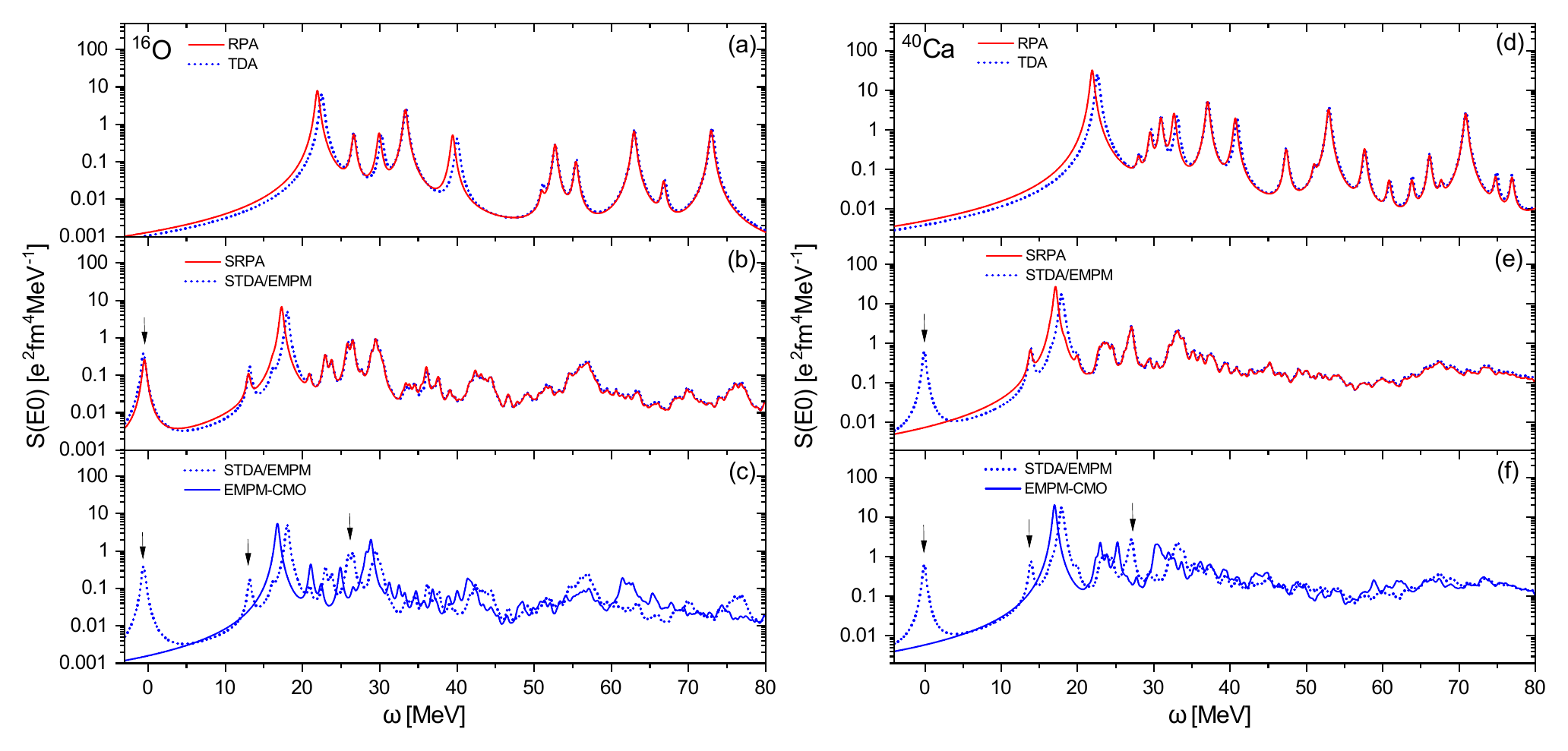}
\caption{(Color online) \label{fig4} Monopole  strength functions  in $^{16}$O and $^{40}$Ca.}
\end{figure*}

\begin{figure*}[ht!]
\includegraphics[width=\textwidth]{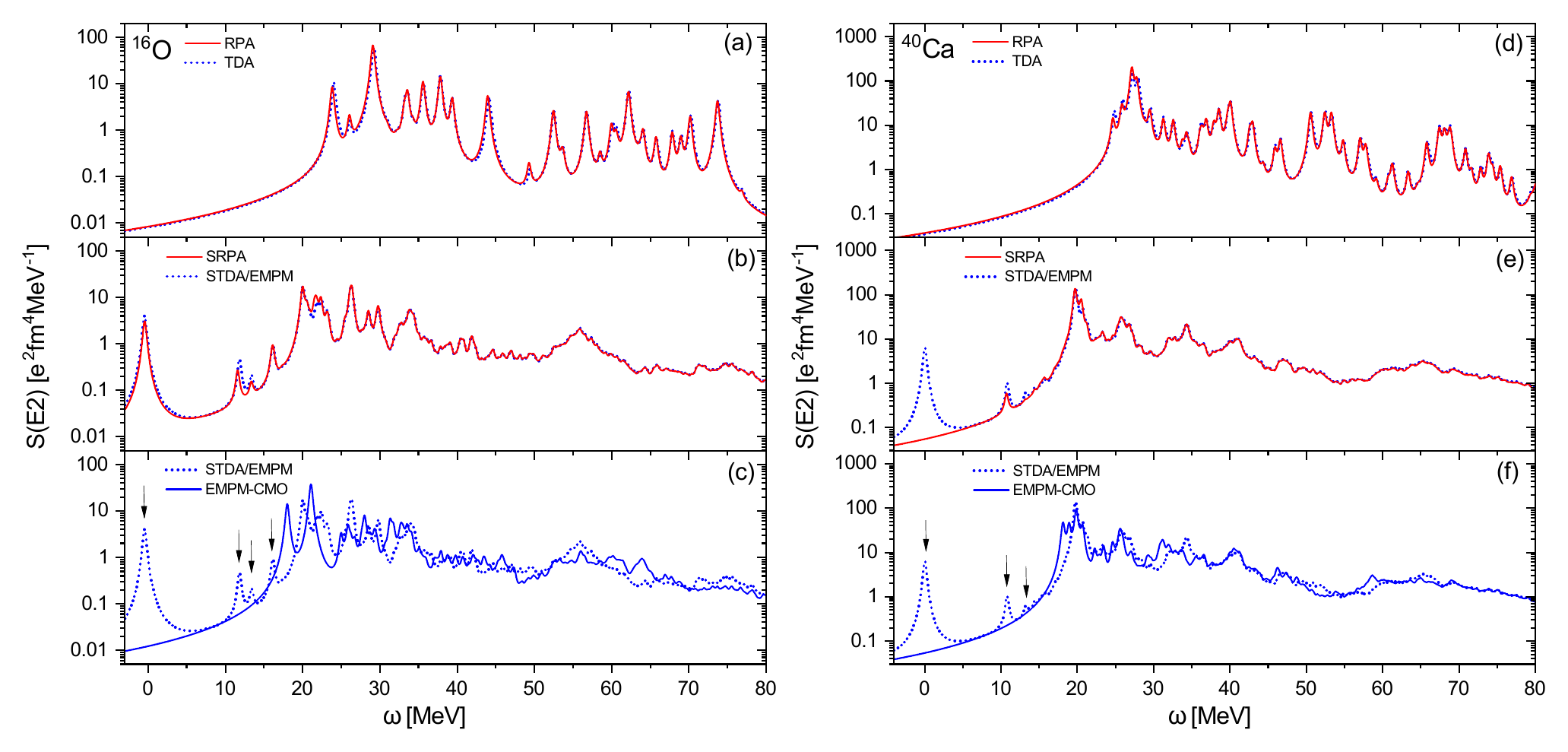}
\caption{(Color online) \label{fig5} Quadrupole strength functions  in $^{16}$O and $^{40}$Ca.}
\end{figure*}

\section{ Short outline of the methods}
\subsection{STDA and SRPA}
The SRPA eigenvalue equations are
\begin{equation*}
\begin{pmatrix}
\cal{A }& \cal{B}\\
-\cal{B}^*& -\cal{A}*
\end{pmatrix} 
\begin{pmatrix}
\cal{X}_\nu \\
\cal{Y}_\nu
\end{pmatrix}
= \omega_\nu
\begin{pmatrix}
\cal{X}_\nu \\
\cal{Y}_\nu
\end{pmatrix}.
\end{equation*} 
Here
\begin{equation*}
\cal{A} =
\begin{pmatrix}
A_{11} & A_{12}\\
A_{21}& A_{22}
\end{pmatrix}, 
\,\,\,
\cal{B}=
\begin{pmatrix}
B_{11} & 0\\
0 & 0
\end{pmatrix},
\end{equation*} 
and
\begin{equation*}
\cal{X}_\nu= 
\begin{pmatrix}
X_\nu ^{(1)}\\
X_\nu ^{(2)}
\end{pmatrix},
\,\,\,\,
\cal{Y}_\nu =
\begin{pmatrix}
Y_\nu ^{(1)}\\
Y_\nu ^{(2)}
\end{pmatrix},
\end{equation*} 
where the  labels 1 and 2 refer to the p-h and the 2p-2h subspaces and 
$X$ and $Y$ are
 the forward and backward amplitudes.
 Concerning the submatrices,
 $A_{11} = \{\bra{i} H \ket{j}\}$ (i=ph, j=p'h') is just the TDA Hamiltonian matrix,
 $A_{22}= \{\bra{ij} H \ket{kl}\}$ are the matrix elements of the Hamiltonian in the 2p-2h subspace, 
 $A_{12} =\{\bra{i}H \ket{jk}\}$ provides the p-h to  2p-2h coupling, and
   $B_{11} = \{\bra{0} H \ket{ij}\}$  is the RPA coupling to the ground state.
The other non diagonal blocks vanish ($B_{12}= B_{21} =B_{22} =0$)  because they are evaluated 
using the HF vacuum instead of the correlated ground state and with only a two-body Hamiltonian~\cite{Yannou87,GGC2011a}.

\begin{figure*}[ht!]
\includegraphics[width=\textwidth]{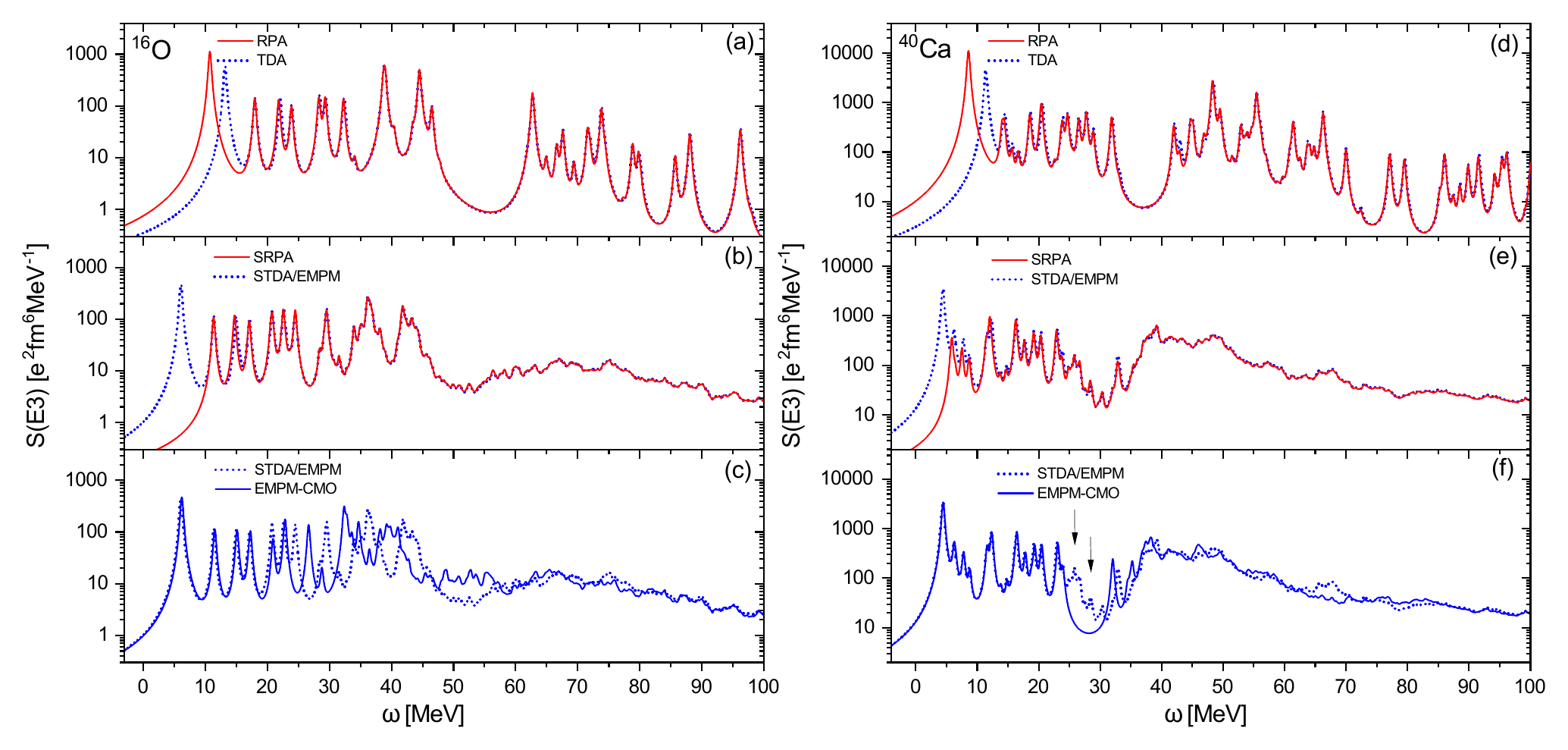}
\caption{(Color online) \label{fig3} $E3$ strength functions  in $^{16}$O and $^{40}$Ca.}
\end{figure*} 
The solution of the above equations yields the eigenvalues $\omega_\nu = E_\nu - E_0$, where $E_0$ is the ground-state energy, $E_\nu$ are energies of the eigenstates
\begin{equation}
 \ket{\Psi_\nu}  = \bigl(O_\nu^{ (1)} + O_\nu ^{(2 )} \bigr) \ket{0},
\end{equation}
where
\begin{eqnarray}
{\cal O}_\nu^{(1)} &=& \sum_{i} \bigl[X_\nu ^{(1)} (i) q^\dagger_1 (i) - Y_\nu ^{(1)} (i) q_1 (i) \bigr] \nonumber\\
{\cal O}_\nu^{(2)} &=& \sum_{ij} \bigl[X_\nu ^{(2)} (ij) q^\dagger_2 (ij) - Y_\nu ^{(2)} (ij)  q_2 (ij) \bigr].
\end{eqnarray}
Here  $q^\dagger_1(q_1)$  and $q^\dagger_2 (q_2)$ create (destroy)  p-h (i)  and 2p-2h  (ij) states, respectively.  
 
If we put $B_{11} =0$ (no ground state correlations), we obtain the STDA equations
\begin{equation}
{\cal A} {\cal X}_\nu =  \omega_\nu {\cal X}_\nu
\end{equation}
whose eigenstates are simply of the form 
\begin{equation}
\ket{\Psi_\nu} = \Bigl[\sum_i X_\nu ^{(1)} (i) q^\dagger_1 (i)+ 
\sum_{ij} X_\nu ^{(2)}(ij) q^\dagger_2 (ij)\Bigr] \ket{0}.
\label{STDAwf}
\end {equation}

\subsection{EMPM}
The EMPM  goes through three steps. We first map the p-h configurations  into  a TDA phonon basis 
\begin{equation}
\{\ket{ph}\} \rightarrow \{\ket{\lambda}\} = \{O^\dagger_\lambda \ket{0}\}.
\end{equation}
Starting from  the TDA one-phonon states $\ket{\alpha_1}= \ket{\lambda}$, we generate iteratively an  orthonormal  basis of $n$-phonon ($n=2,3..$) correlated states $\ket{\alpha_{n}}$ through various steps.
Assuming known the  $(n-1)$-phonon basis states $\ket{\alpha_{n-1}}$, we construct   $n$-phonon states
\begin{equation}
 \label{Olam}
 \ket{i} = \ket{\lambda \alpha_{n-1}}=  O^\dagger_\lambda \ket{\alpha_{n-1}}.
 \end{equation}
From this redundant set we extract a basis of linearly independent (but not orthogonal) states through the Cholesky decomposition method and formulate, in  the basis so obtained, the generalized eigenvalue equation within the $n$-phonon subspace 
\begin{equation}
\label{eig1}
\sum_{jk} \bigl({\cal A}^{(n)}_{ik} - E^{(n)} \delta_{ik} \bigr)   {\cal D}^{(n)}_{kj} C^{(n)}_{j}  = 0.
\end{equation}
In Eq. (\ref{eig1})
\begin{equation}
\label{D}
 {\cal D}^{(n)}_{ij}= \langle i \mid j \rangle
 \end{equation}
is the overlap or metric matrix and
\begin{eqnarray}
 {\cal A}^{(n)}_{ ij} =     E_i \delta_{ij} + {\cal V}^{(n)}_{ij}, 
\label{A}
\end{eqnarray}
where $E_i$ is the unperturbed energy of the $n$-phonon state (\ref{Olam}).
The formulas giving  the overlap matrix ${\cal D}$ and the phonon-phonon interaction  ${\cal V}$  can be found, for instance, in Ref. \cite{Bianco12}.

The solution of Eq. (\ref{eig1}) yields an orthonormal basis  of states
\begin{equation}
\ket{\alpha_n} =\sum_{\lambda \alpha_{(n-1)} } C^{\alpha_n}_{\lambda \alpha_{(n-1)} } \ket{\lambda \alpha_{(n-1)} }
\label{alphan}
\end{equation}
within the $n$-phonon subspace. 
 The iteration of the procedure up to an arbitrary $n$ produces a set of states which,  added to   HF ($\ket{0}$) and TDA   ($\{\ket{\alpha_1}\}= \{\ket{\lambda}\}$), form an orthonormal basis $\{ \ket{\alpha_n}\}$ ($n=0,1,2,3,...$) with energies $E_{\alpha_n}$, spanning the full multiphonon space.
 
In such a space, we solve the final eigenvalue equations
\begin{equation}
\label{eigfull}
\sum_{ \alpha_n \beta_{n'}} \Bigl((E_{\alpha_n} - {\cal E}_\nu) \delta_{\alpha_n \beta_{n'}}  +   {\cal V}_{\alpha_n \beta_{n'}} \Bigr){\cal C}^{\nu}_{\beta_{n'}} = 0,
\end{equation}
where ${\cal V}_{\alpha_n \beta_{n'}}  = 0$ for  $n' = n$. The formulas giving  ${\cal V}_{\alpha_n \beta_{n'}} $ ($n' \neq n$) can be found in Ref.  \cite{DeGreg2022}.
 The resulting eigenvectors (with corresponding energies ${\cal E}_\nu$) 
\begin{equation}
\label{Psifull}
\ket{\Psi_\nu} = \sum_{n,\alpha_n}  {\cal C}_{\alpha_n}^\nu \ket{\alpha_n},
\end{equation}
including the ground state $\ket{\Psi_0}$, are fully correlated. 
In order to make a comparison with the other two approaches
we consider a space including up to two phonons and neglect the coupling between the HF vacuum and the two-phonon states ($\bra{\alpha_2}H \ket{0} =0$) not considered in STDA and SRPA.
Under this assumption the ground state is simply the HF state $\ket{0}$ and the excited states have the structure
\begin{equation}
\label{Psi2}
\ket{\Psi_\nu} =\sum_\lambda {\cal C}_{\lambda}^\nu \ket{\lambda} + \sum_{\lambda_1 \lambda_2} {\cal C}_{\lambda_1 \lambda_2}^\nu \ket{\lambda_1 \lambda_2}
\end{equation}
where we made use  of Eq. (\ref{alphan}).

\begin{figure}[ht!]
\includegraphics[width=\columnwidth]{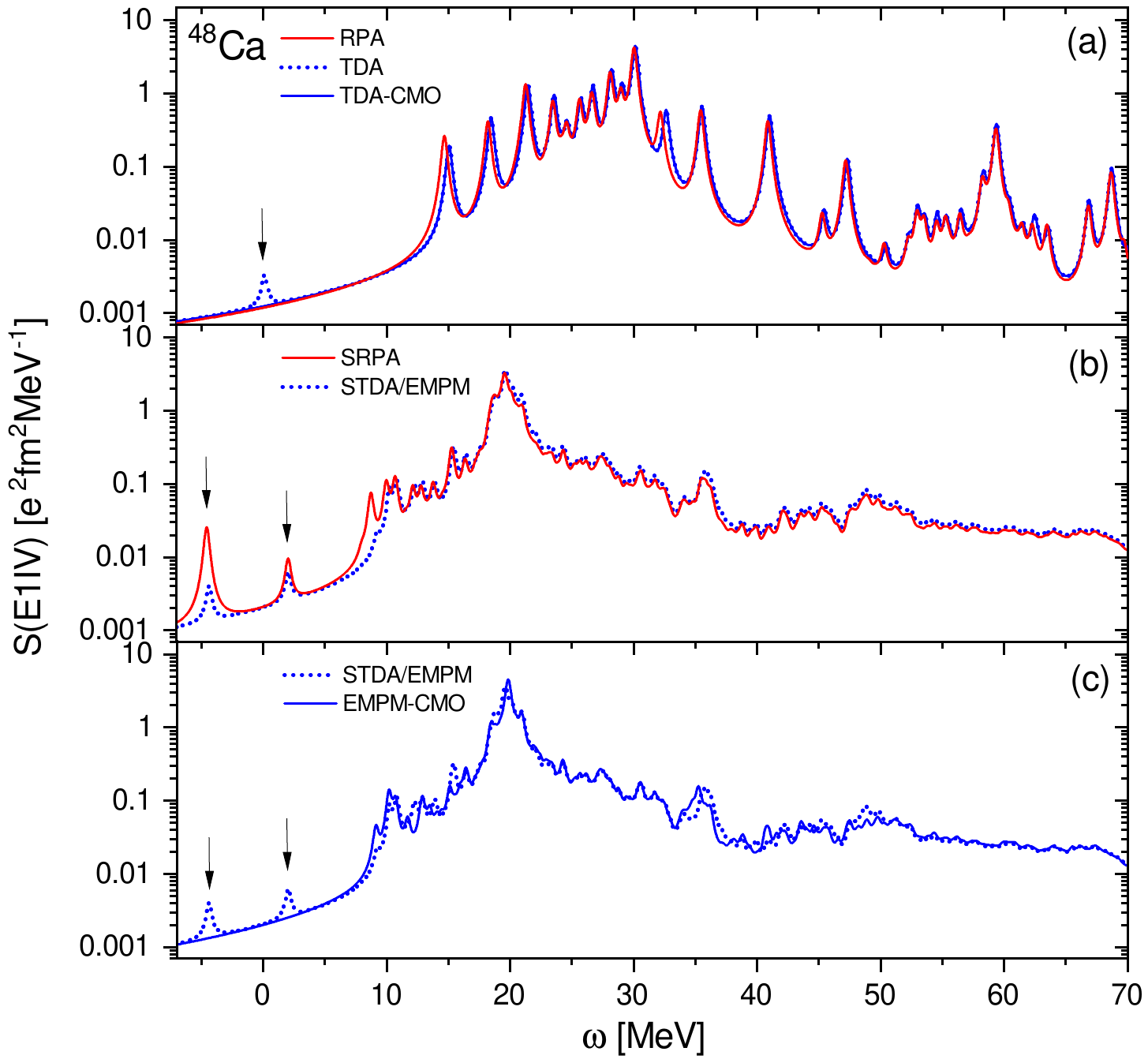}
\caption{(Color online) \label{fig6} Isovector $E1$ strength functions  in $^{48}$Ca.}
\end{figure}

\subsection{Comparative analysis}
We have seen already that SRPA turns into STDA if one neglects the ground state correlations which in SRPA are treated in the quasi-boson approximation.
Within the p-h + 2p-2h space, the EMPM wavefunctions (\ref{Psi2}) can assume the structure of the STDA states (\ref{STDAwf})  by expressing the TDA phonons $\lambda$ in terms of the p-h configurations. Since there is a one to one correspondence between p-h and TDA states, the EMPM is completely equivalent to STDA. We will demonstrate it on many numerical examples in the next section. 

With respect to both STDA and SRPA, the EMPM adopts a correlated basis which can be safely truncated. Moreover, it allows naturally the extension of the calculations  beyond 2p-2h in spaces  including three-phonon and, even, four-phonon  states and yields an explicitly correlated ground state without resorting to any approximation apart from a space truncation.  

The EMPM has the disadvantage that it has to deal with a redundant basis which renders the procedure more involved. On the other hand, just the use of such a basis allows a complete and exact elimination of the spurious admixtures induced by the c.m. motion for any single-particle basis. 
In fact, we can first generate a basis of c.m. free TDA states orthogonal to the c.m.  spurious state $\ket{\lambda_{c.m.}}$ by applying the Gram-Schmidt  orthogonalization to the p-h configurations  \cite{Bianco14}. The SVD allows us to extend the orthogonalization procedure to all $n$-phonon subspace \cite{DEGREGORIO2021}. For $n=2$, for instance, we distinguish
the set of spurious states  $\{\ket{i_{c.m.}}\} = \{\ket{\lambda \lambda_{c.m.}},    \ket{\lambda_{c.m.} \lambda_{c.m.}} \}$ from the other two-phonon states $\{\ket{j}\} = \{\ket{\lambda \lambda'} \}$.

\begin{figure}[ht!]
\includegraphics[width=\columnwidth]{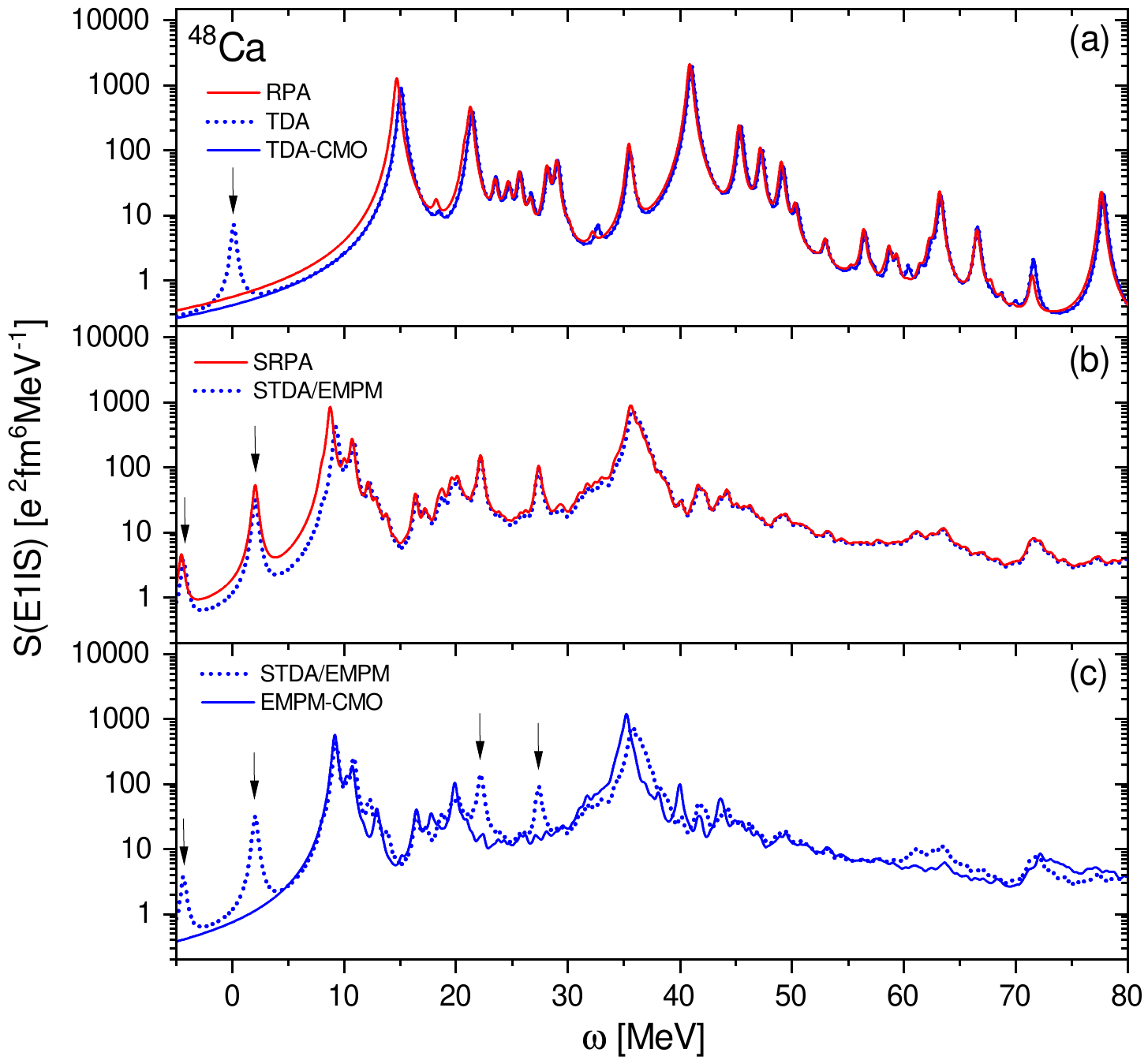}
\caption{(Color online) \label{fig7} Isoscalar $E1$ strength functions  in $^{48}$Ca.}
\end{figure}

The SVD method decomposes the rectangular overlap matrix 
\begin{equation}
{\cal D}_{j ,i_{c.m.} } = \langle i_{c.m.} \mid j \rangle
\end{equation}
into two mutually orthogonal diagonal blocks defining two subspaces. One is spanned by the c.m. free basis $\{\ket {\alpha}\}$, the other  by the c.m. spurious states $\{\ket{\alpha_{cm}} \}$. The two subspaces are mutually orthogonal
\begin{equation}
 \langle \alpha_{cm} \mid  \alpha \rangle =0. 
 \end{equation}
The procedure can be extended to any $n$-phonon subspace ($n=3,4,..$).

We will investigate how the differences between the three approaches and, especially, the different treatment of the c.m. motion have a quantitative impact on the nuclear multipole responses. To our knowledge, only in RPA are the physical excited states automatically decoupled from the spurious c.m. motion.

\section{Numerical implementation and results}
We adopt an intrinsic  Hamiltonian of the form 
\begin{equation}
H = T_{int} + V = \sum_{i<j} \Bigl( \frac{\mathbf{p}^2_{ij}}{2A m} + V_{ij} \Bigr)
\label{Eq:Ham} 
\end{equation}
where $\bold{p}_{ij} = \mathbf{p}_i - \mathbf{p}_j$ and  $m$ is the nucleon mass, for both protons and neutrons.
We adopt the  UCOM  potential \cite{Feldmeier1998} to generate a HF basis from a HO model space truncated in the major oscillator quantum number N$_{max}$ =6 and the oscillator length $b=1.7$ fm. Being derived from Argonne V18 \cite{Argonne18}, UCOM can be considered a realistic two-body 
potential which avoids the  double-counting  problems affecting entirely phenomenological potentials~\cite{Tsel2007,Tsel2013}.

\begin{figure}[t!]
\includegraphics[width=\columnwidth]{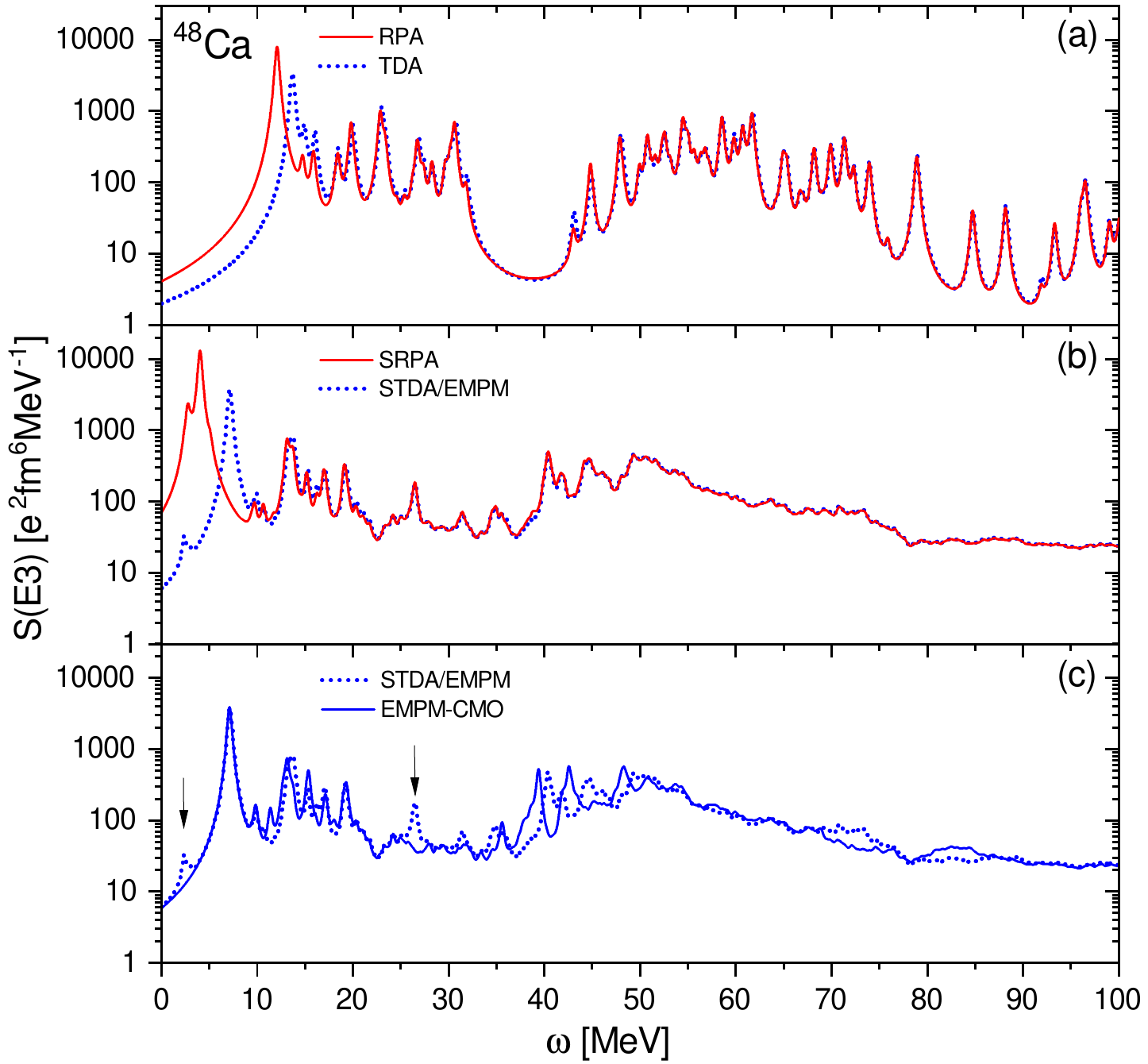}
\caption{(Color online) \label{fig8} $E3$ strength functions  in $^{48}$Ca.}
\end{figure}

We will compute the strength functions of the electric multipole operators  whose general form is
\begin{equation}
{\cal M}(E; \lambda \mu)= \sum_k e_k r^{(\lambda+ n)}_k  Y_{\lambda \mu} (\hat{k}),
\label{MEL}
\end{equation}
where, unless specified otherwise, we use the bare charges, $e_k =e$ for protons and $e_k =0$ for neutrons.
The strength function is given
\begin{equation}
S(E\lambda,\omega) = \sum_\nu B(E\lambda, \Psi_0 \rightarrow \Psi_\nu ) \delta(\omega - \omega_\nu),
\end{equation}
where $\omega_\nu =E_\nu - E_0$ and 
\begin{equation}
B(E\lambda, \Psi_0 \rightarrow \Psi_\nu ) = | \langle \Psi_\lambda \parallel {\cal M}(E\lambda)\parallel 0 \rangle |^2
\end{equation}
is the reduced transition probability.
We will replace the delta function with a Lorentzian 
\begin{equation}
\delta(\omega - \omega_\nu)  \rightarrow \frac{1}{2\pi} \frac{\Gamma}{(\omega - \omega_\nu)^2 +\Gamma^2/4}
\end{equation}
of width $\Gamma = 0.5 $~MeV for presentation purposes.



The choice of the intrinsic Hamiltonian, Eq.~(\ref{Eq:Ham}), specifically, of the intrinsic two-body operator for the kinetic energy, is relevant in what follows.  
To some extent, the c.m. kinetic energy can be subtracted by employing the same form as the total kinetic-energy operator, which is a single-particle
operator, but with a correction to the nucleon mass, 
\begin{equation} 
T_{int}^{(1)}= \left( 1 - \frac{1}{A} \right)\sum_{i=1}^A \frac{\mathbf{p}^2_{i}}{2m}. 
\label{Eq:Tint1} 
\end{equation} 
This prescription is often used in RPA calculations based on phenomenological functionals \cite{ColoCPC2013}. However, with such choice, the formal conditions for the spurious state to appear at zero energy are not met. Let us consider, for example, the spurious c.m. motion operator $O_{sp}=\frac{1}{A}\sum_{i=1}^A \vec{r}_i$, which contaminates the dipole channel. As pointed out also in Ref.~\cite{Papa14a} for the total kinetic energy, $T_{int}^{(1)}$ does not commute with $O_{sp}$, so the total energy weighted sum rule does not vanish. As a result, all RPA calculations employing $T_{int}^{(1)}$ produce a spurious state at an energy of finite value (real or imaginary), regardless of how large the p-h-space is. By contrast, $T_{int}$ does commute with $O_{sp}$ and RPA implementations employing it can produce a spurious state at practically zero energy (in terms of the numerical precision of the overall implementation). As a demonstration, we compare in Fig.~\ref{fig:spurious} RPA results obtained with $T_{int}$, {\em i.e.}, by employing the Hamiltonian of Eq.~(\ref{Eq:Ham}), and with $T_{int}^{(1)}$, Eq.~(\ref{Eq:Tint1}).

Specifically, we show the RPA response of $^{40}$Ca to the isoscalar dipole operator given by Eq.~(\ref{MEL}) with $\lambda=1,n=2,e_p=e_n=e$.  
In the case of $T_{int}$, even in such a small basis ($N_{max}=6$ or 60 p-h states) the spurious state occurs at 0.09~MeV. 
In the case of $T_{int}^{(1)}$, it appears at 0.86~MeV. 
The respective values in a basis of $N_{max}=10$ are 0.006 and 0.322~MeV.
All eigenstates are affected by the choice of kinetic-energy operator and the same holds in all channels as exemplified in the quadrupole case, also shown in Fig.~\ref{fig:spurious}.   

\subsection{Nuclear response in $^{16}$O and $^{40}$Ca}

 \subsubsection{Isovector electric dipole response}
For the isovector electric dipole operator, $n=0$ and $\lambda=1$ in Eq. (\ref{MEL}),
we replace the bare charges with
the effective ones $e_k = (N/A) e$ for protons and $e_k = - (Z/A) e$ for neutrons in order to minimize the impact of the c.m. coordinates. 
They are obtained by referring the nucleonic coordinates 
to the c.m. coordinate, $\vec{r}_k \rightarrow (\vec{r}_k - \vec{R}_{c.m.})$.  
Such a replacement would be unnecessary within the EMPM. In fact, after the SVD treatment,  we obtain the same strengths whether  we use bare or  effective charges.

The behavior of $E1$ strength functions is very similar in both  $^{16}$O and $^{40}$Ca (Fig. \ref{fig1}).
As shown in panels (a) and (d),  the use of the effective charges ensures the complete removal of the c.m. admixtures in RPA and, to a very large extent, in TDA. In the latter case,
the residual impurity is removed after the implementation of the Gram-Schmidt orthogonalization procedure.  Remarkably enough, TDA and  RPA strength functions  are practically
indistinguishable. 

 Panels (b) and (e) show that the STDA  strength distribution is identical to the one computed within the EMPM before the implementation of the SVD. Both strengths are nearly indistinguishable from the  one obtained in SRPA. 
 
 The implementation of the SVD method has a visible effect (panels (c) and (f)). It identifies and removes three spurious peaks, present in all approaches, one at negative energy and two at high energy.  However, it does not alter significantly the overall profile of the strength, especially in the region of the giant dipole resonance.

\begin{figure}[t!]
\includegraphics[width=\columnwidth]{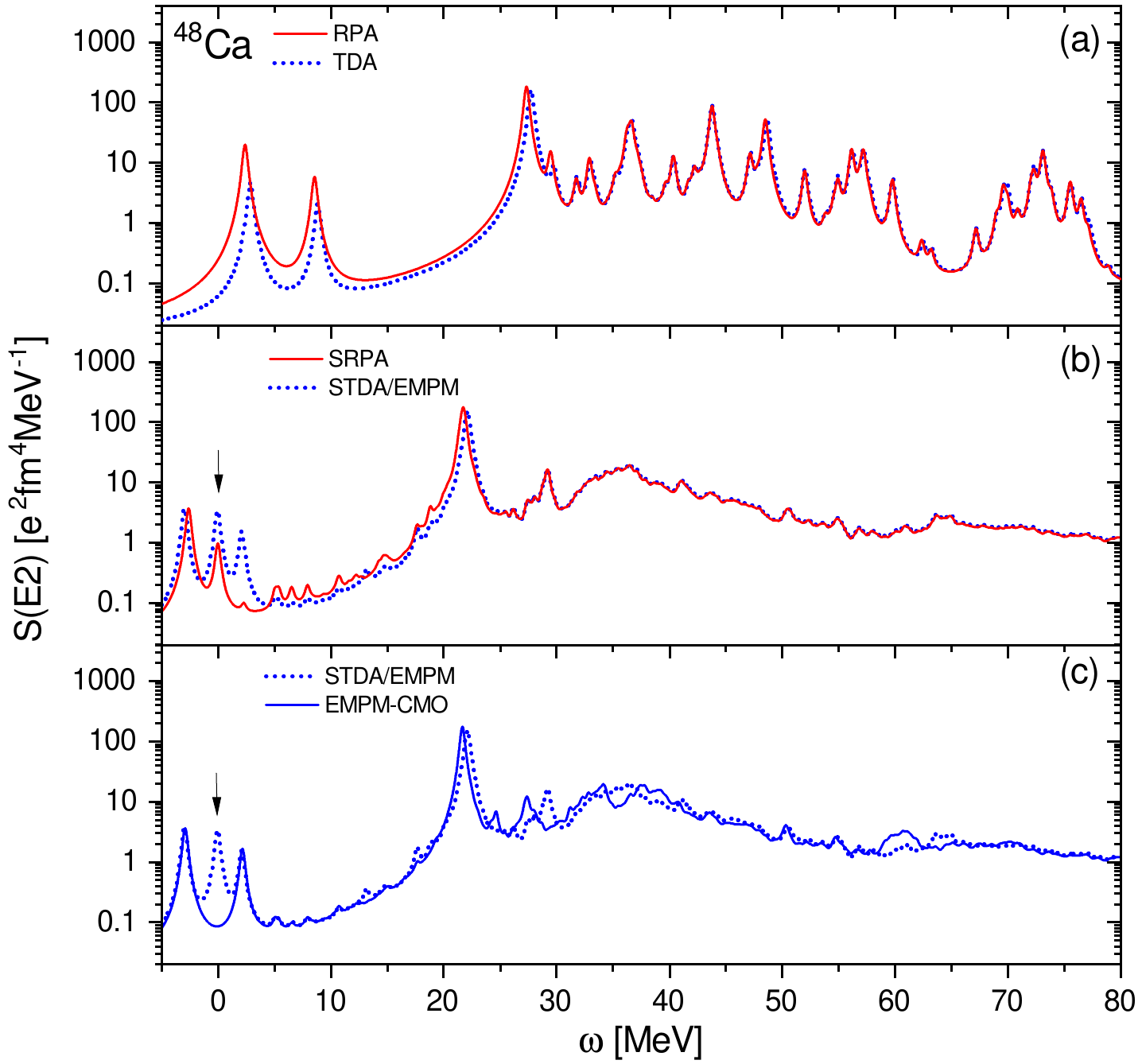}
\caption{(Color online) \label{fig9}Quadrupole strength function  in $^{48}$Ca.}
\end{figure}

\subsubsection{Isoscalar dipole response}
For the isoscalar dipole response we use the operator 
\begin{equation}
{\cal M}_{IS} (E; 1 \mu)=  e \sum_k (r_k^2 - \frac{5}{3}  \langle r^2 \rangle )r_k Y_{1 \mu} (\hat{k}).
\label{E1IS}
\end{equation} 
The linear term is introduced in order to minimize the spurious contributions coming from the c.m. motion.
The strength function for $^{16}$O and for $^{40}$Ca is shown in Fig. \ref{fig2}. In analogy with the case of the isovector $E1$ transitions, the TDA and STDA isoscalar $E1$ strength distributions overlap to a very large extent with the RPA and SRPA corresponding distributions, respectively. 
Despite the inclusion of the linear term in Eq. (\ref{E1IS}), a spurious peak occurs at zero energy in RPA and, especially, in TDA (panels (a) and (d)). In TDA, it disappears after the Gram-Schmidt orthogonalization.  As we move to STDA and SRPA (panels (b) and (e)), we observe that the low-lying spurious peak remains and drops to negative energy.   The implementation of the SVD method not only eliminates such a peak but reveals and eliminates two additional spurious peaks.  More in general, it shows that, if not eliminated, the spuriousness spreads over fairly large energy intervals and alters  portions of the profile of the strength distribution (panels (c) and (f)).

\begin{figure}[t!]
\includegraphics[width=\columnwidth]{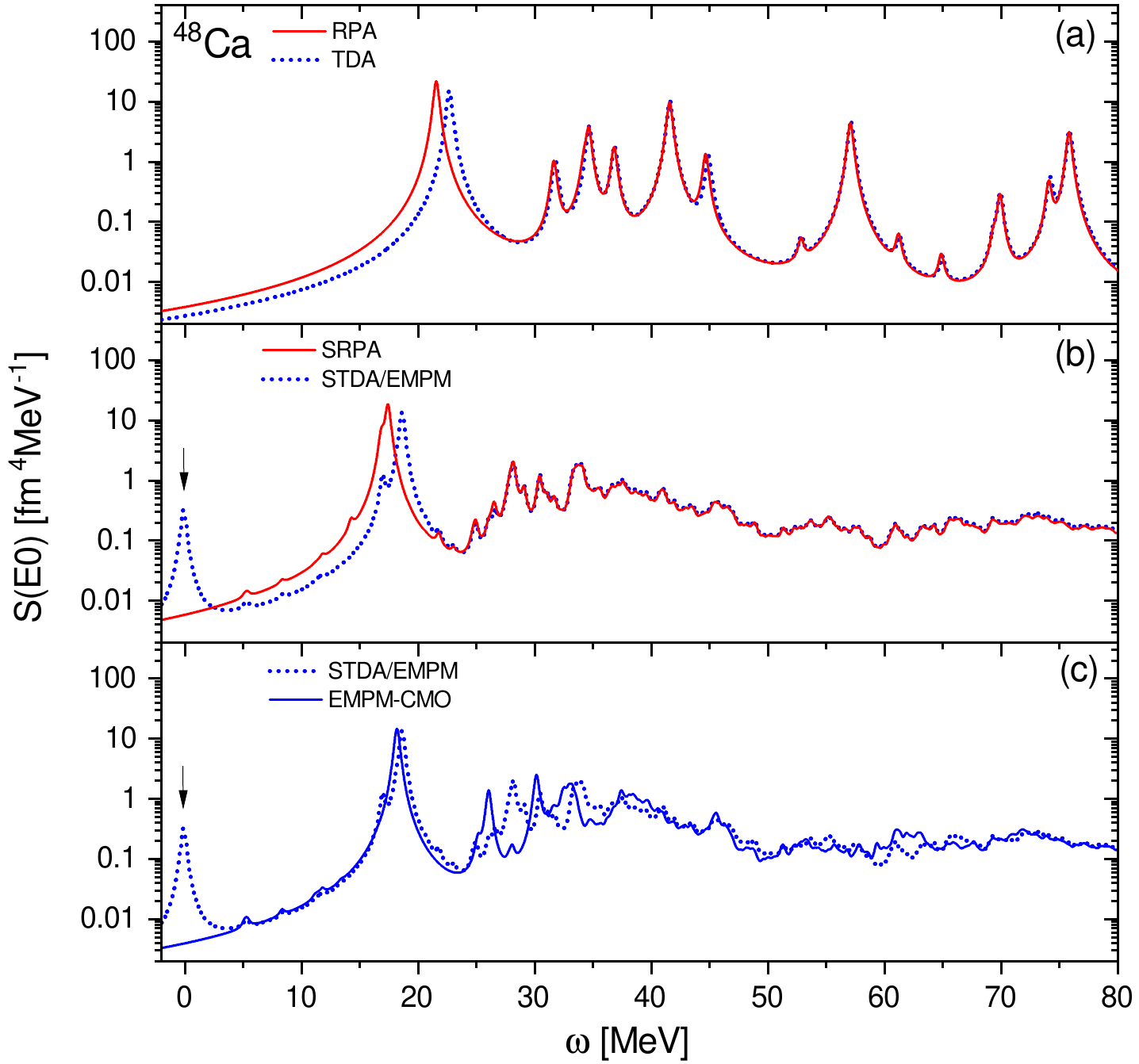}
\caption{(Color online) \label{fig10}Monopole strength function  in $^{48}$Ca.}
\end{figure} 
\subsubsection{Octupole response}
The TDA and RPA octupole ($n=0$ and $\lambda =3$ in Eq. (\ref{MEL})) strength distributions overlap over a large interval at high energy but  differ considerably from each other in the low energy sector (Fig. \ref{fig3} panels (a) and (d)). Both approaches yield a strong low energy peak. However, the one obtained in RPA is more than 5 MeV below in energy and disappears in SRPA (Fig. \ref{fig3} panels (b) and (e)) because it is obtained at imaginary energy. 
In STDA  (EMPM) such a low-lying peak is still present  even after the implementation of the SVD method (Fig. \ref{fig3} panels (c) and (f)). Therefore, it corresponds to a genuine physical resonance. The c.m. affects the spectrum only at high energy, above 20 MeV.  SVD disposes of all spurious peaks as well as of the residual contaminations.


\begin{figure*}[t!]
\includegraphics[width=16cm]{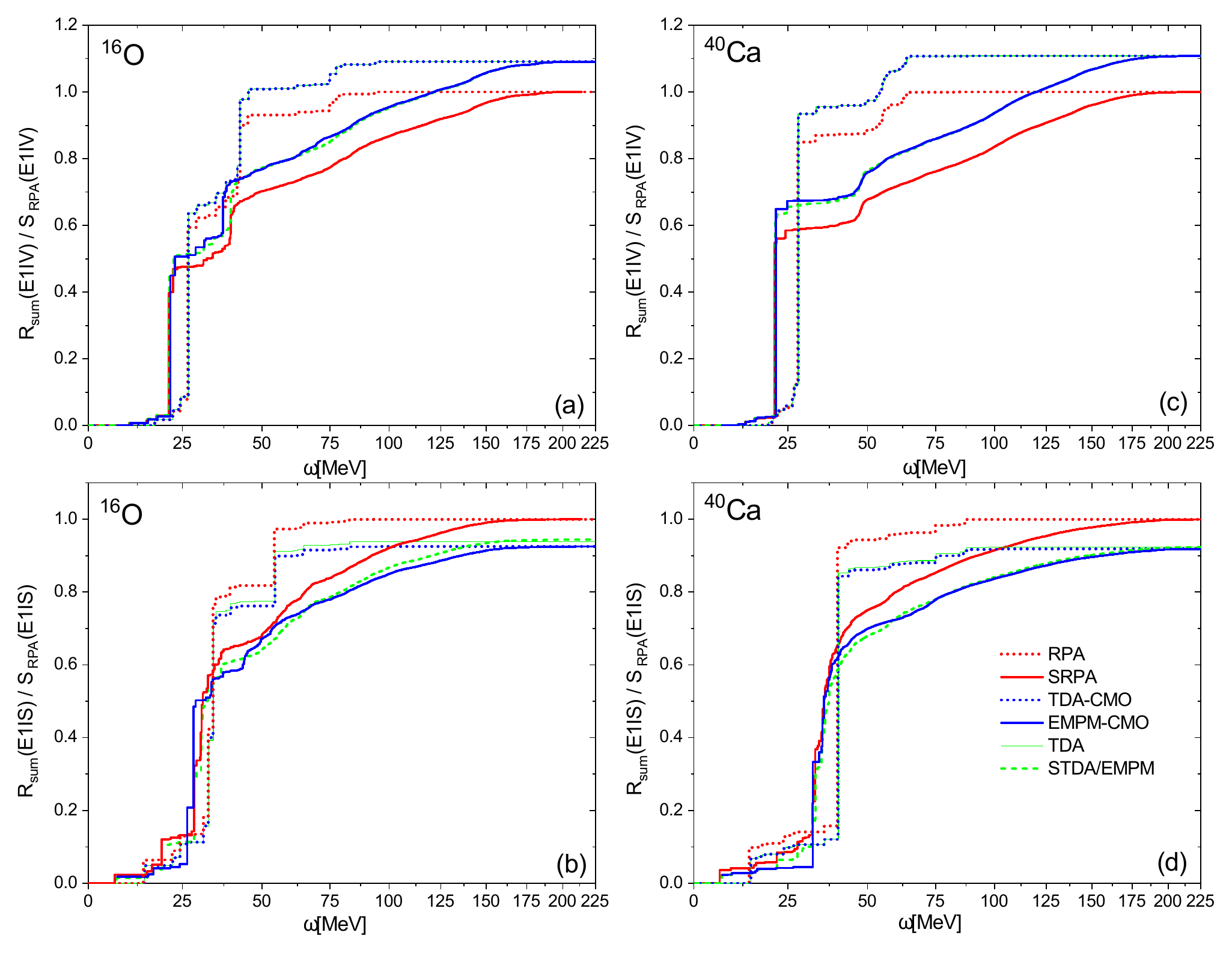}
\caption{(Color online) \label{fig11} Isovector and isoscalar $E1$ energy weighted running sums in $^{16}$O and $^{40}$Ca.}
\end{figure*}

 \subsubsection{Monopole and Quadrupole responses}
From  comparing  Fig.\ref{fig4} and Fig.\ref{fig5} we observe that monopole ($n=2$ and $\lambda =0$) in Eq. (\ref{MEL}) and quadrupole ($n=0$  and $\lambda =2$) 
responses exhibit similar characteristics in  $^{16}$O.
We notice a near overlap between TDA and RPA (panel (a)) as well as between STDA and SRPA (panel (b)) strength distributions, especially in the quadrupole case. Humps at zero energy appears in both STDA and SRPA.  Such peaks are induced by the c.m. motion through $[1^-\otimes 1^-]^{0^+,2^+}$ coupling of the spurious state with itself. 
The c.m. spuriousness is not concentrated only around zero energy but spreads over the whole energy interval. It generates several spurious peaks, in addition to the zero energy one, and contaminates other transitions. All spurious peaks as well the contaminations of other transitions are eliminated once the SVD method is implemented.

Fig. \ref{fig4} and Fig. \ref{fig5} show that the features of the monopole and quadrupole responses  in $^{40}$Ca  are very similar to those observed in $^{16}$O. 
The only significant discordance is that in $^{40}$Ca, the zero energy spurious monopole and quadrupole peaks appear in STDA but not in SRPA. The reason is that the SRPA energy eigenvalues corresponding to these peaks are imaginary. Implementation of the SVD method removes not only the zero energy bumps but also all states with spurious admixtures.

\begin{figure*}[t!]
\includegraphics[width=16cm]{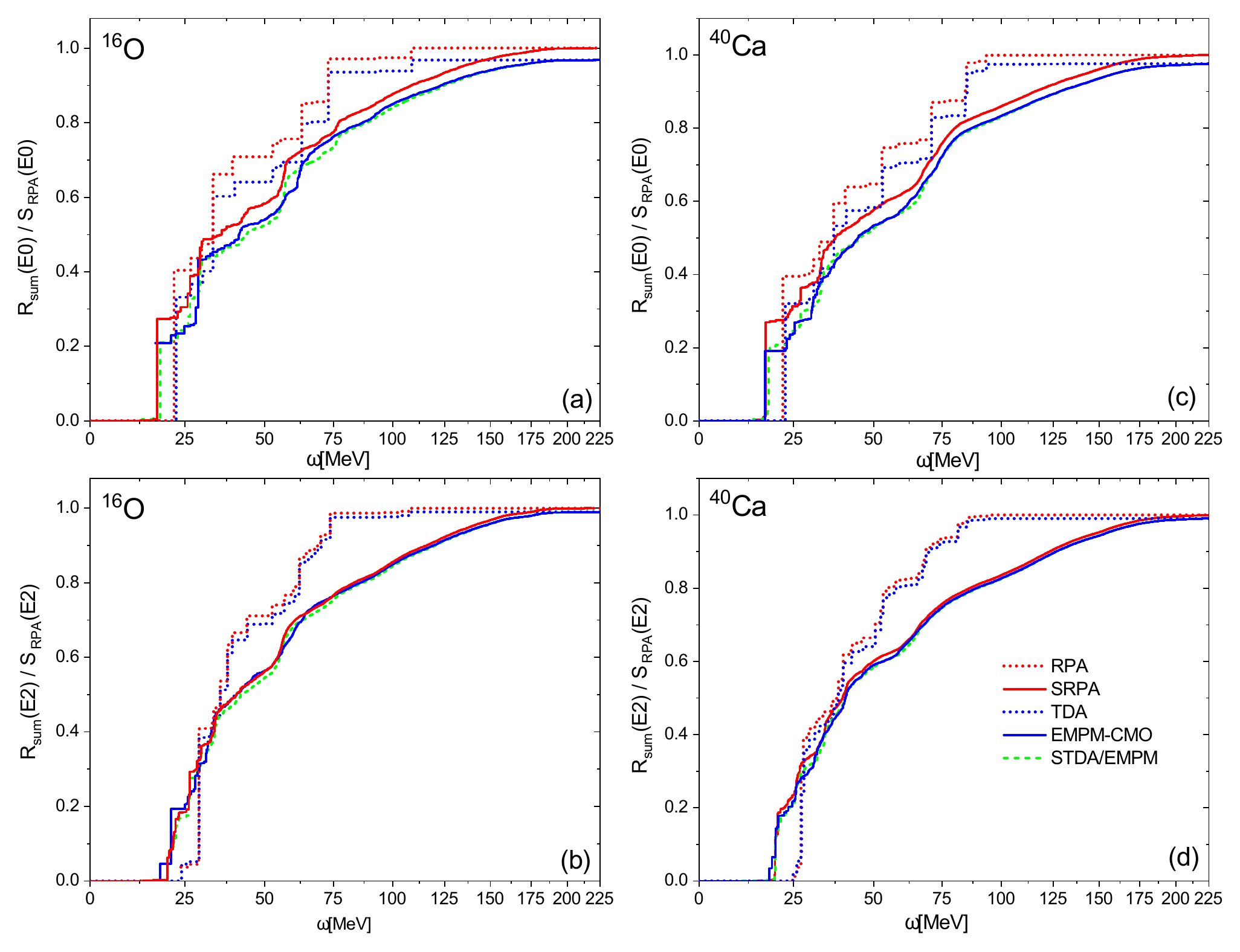}
\caption{(Color online) \label{fig12} $E0$ and $E2$ energy weighted running sums in $^{16}$O and $^{40}$Ca.}
\end{figure*}

\begin{figure*}[ht!]
\includegraphics[width=16cm]{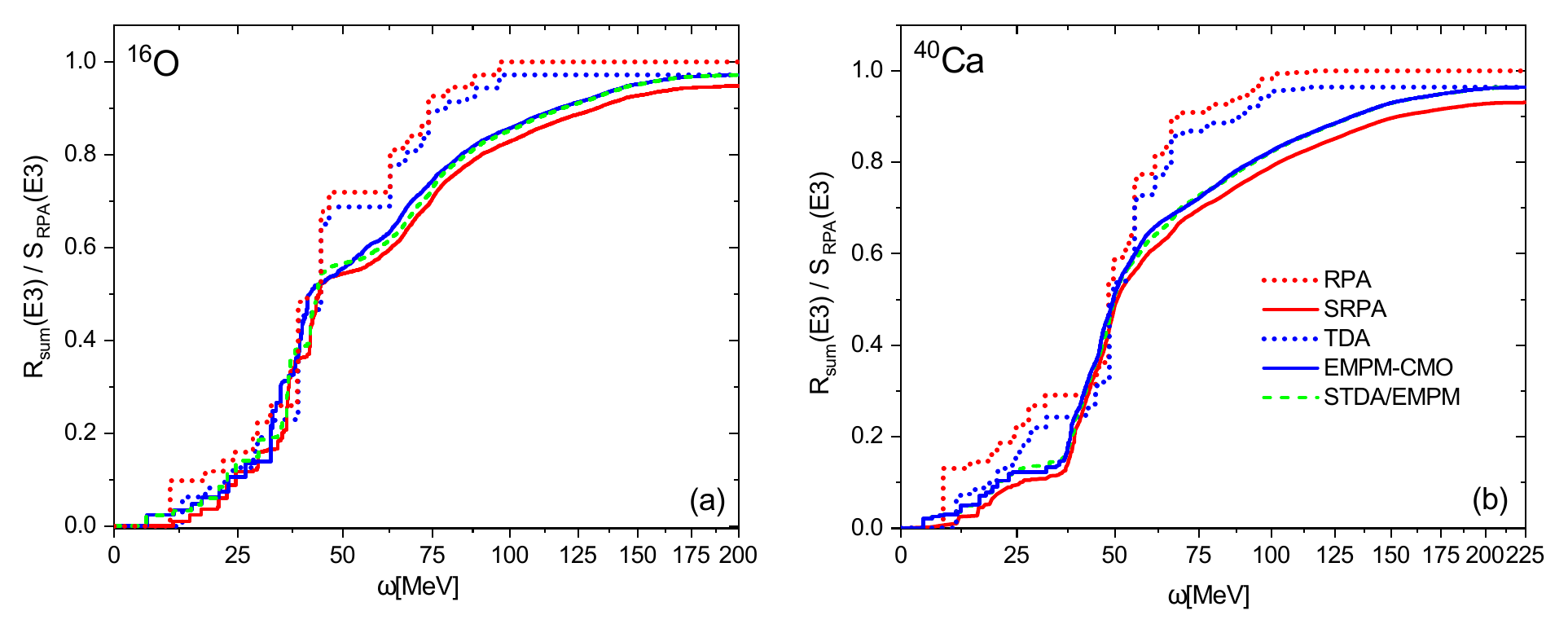}
\caption{(Color online) \label{fig13} $E3$ energy weighted running sums in $^{16}$O and $^{40}$Ca.}
\end{figure*}

\begin{figure*}[ht!]
\includegraphics[width=16cm]{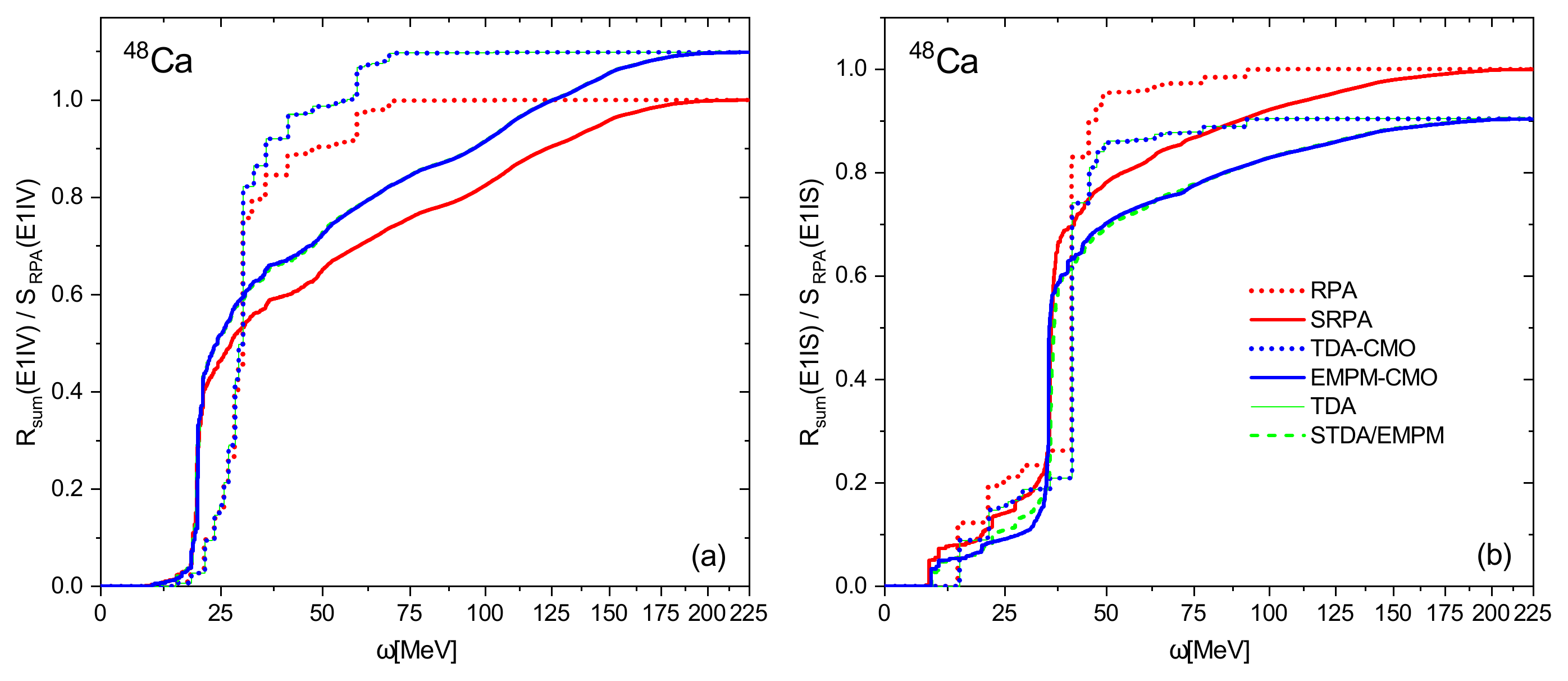}
\caption{(Color online) \label{fig14} Isovector and isoscalar $E1$ energy weighted running sums in $^{48}$Ca.}
\end{figure*}

\begin{figure*}[ht!]
\includegraphics[width=16cm]{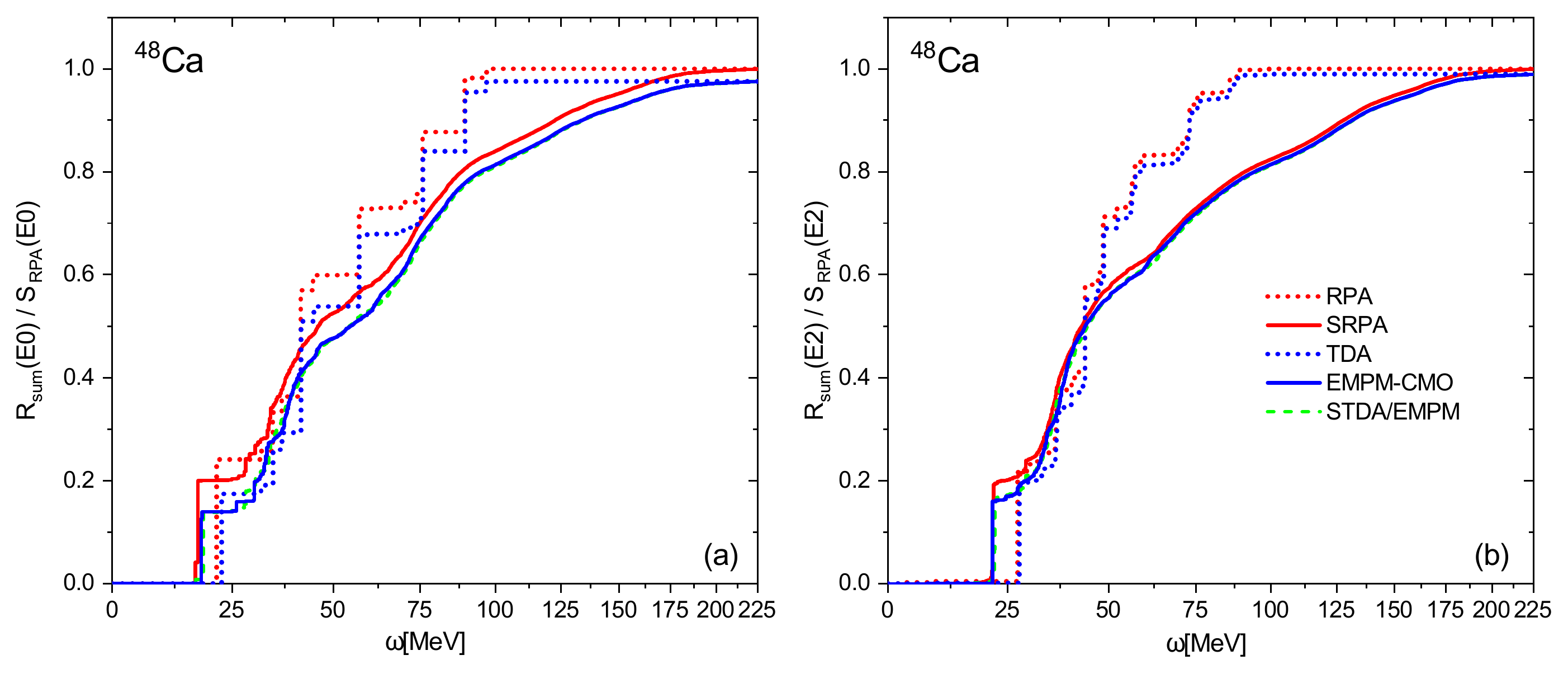}
\caption{(Color online) \label{fig15} $E0$ and $E2$ energy weighted running sums in $^{48}$Ca.}
\end{figure*}
 
\subsection{Nuclear response in $^{48}$Ca}
We examine  $^{48}$Ca separately from  $^{16}$O and  $^{40}$Ca as a qualitatively different system. Specifically, it is not only isospin asymmetric, but it also develops a low-lying quadrupole vibration which can couple to other phonons and affect all channels. 

As shown in Fig. \ref{fig6} , the isovector $E1$ response is very similar in both TDA and RPA as well as in STDA and SRPA.  The c.m.  induces a small peak in TDA and two weak transitions in STDA and SRPA.
Its overall impact on the response is modest as shown in panel (c).

More pronounced is its effect on the isoscalar $E1$ strength function (Fig. \ref{fig7}). The c.m. motion generates several peaks and contaminates several transitions. All these impurities disappear after the SVD treatment.

TDA and RPA as well as STDA and SRPA have a similar $E3$ response at high energy (Fig. \ref{fig8}). At low energy, instead, important deviations are observed.  The RPA low-energy peak is shifted  by about 5 MeV with respect to TDA. In SRPA it is further shifted with respect  to STDA. The SVD method removes a low-energy hump predicted by both STDA and SRPA and washes the residual spurious admixtures, which distort the spectrum in both the low- and high-energy part.

The monopole (Fig. \ref{fig9}) and quadrupole (Fig. \ref{fig10}) responses are similar in the high energy sector but behave differently at low energy. Both TDA and RPA yield two $E2$ low-energy peaks, while the monopole spectra are flat. A  spurious monopole peak at zero energy appears in STDA (EMPM) but not in SRPA, again because of the imaginary nature of the corresponding energy eigenvalue. The discrepancy is solved once such a spurious excitation is removed through the SVD method.
Three low energy quadrupole peaks are generated  in  STDA (EMPM) and two in SRPA. Two of the three peaks survive even after the implementation of SVD.  They are genuine physical states.

However, the excitation energy of one of the two peaks is negative, in both STDA and SRPA, implying that the excited $2^+$ state lies below the HF  state.    
Such an anomaly indicates that, at least for the potential adopted here, it is not appropriate to consider the unperturbed HF as ground  state, a tacit assumption made in RPA, SRPA, and STDA. We need to replace HF with a correlated ground state, but such a replacement would require the inclusion of 3p-3h, or 3-phonon basis states. Such a task is beyond the scope of the present work.

\begin{figure}[ht!]
\includegraphics[width=\columnwidth]{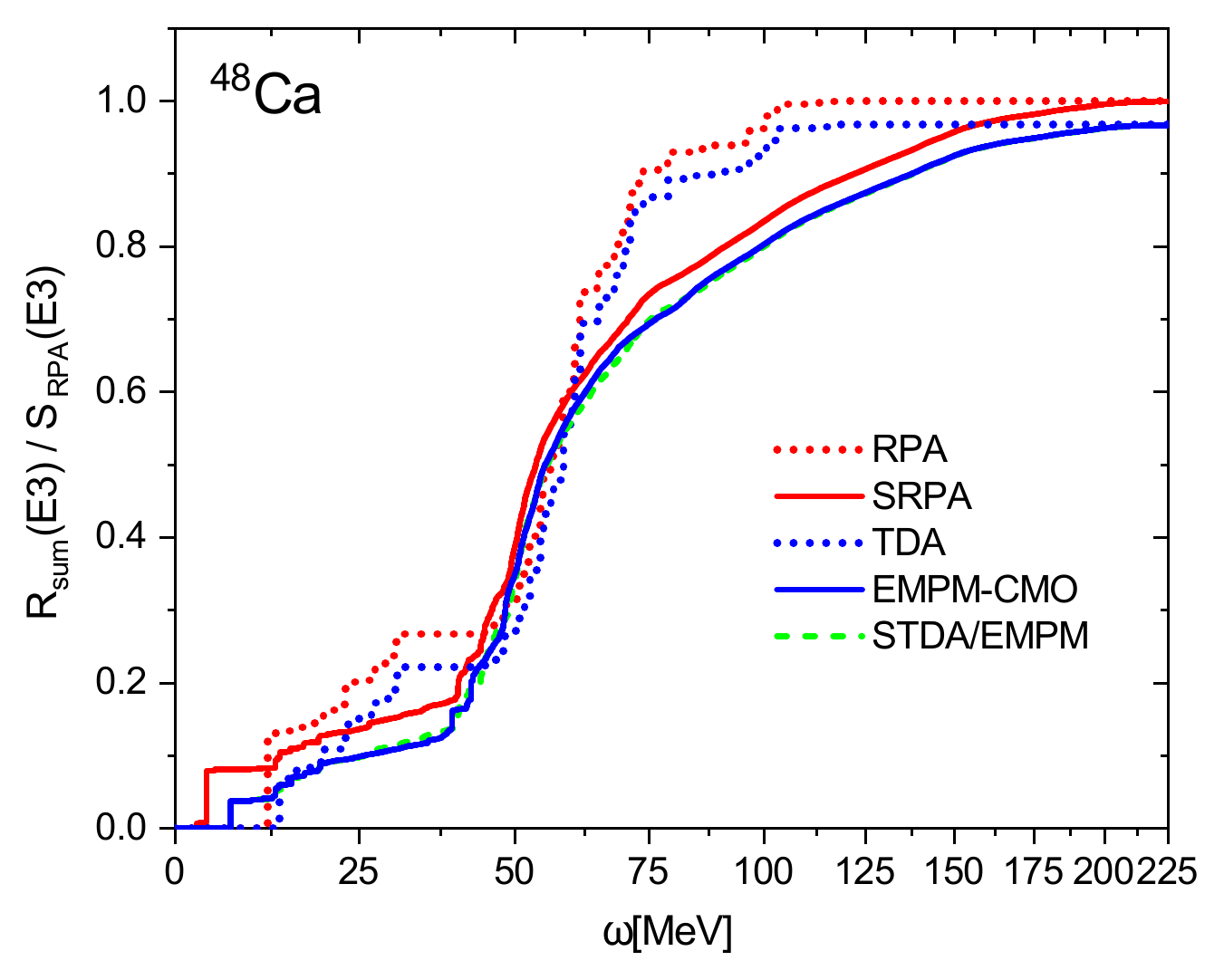}
\caption{(Color online) \label{fig16} $E3$ energy weighted running sums in $^{48}$Ca.}
\end{figure}

\subsection{Running sum}
It is useful to analyze the nuclear response from the perspective offered by the running sum.
Several interesting aspects emerge from examining the plots shown in Figs.  \ref{fig11}, \ref{fig12}, \ref{fig13}, \ref{fig14}, \ref{fig15} and \ref{fig16}. 
 In all nuclei under investigation,  the  running sum  of all transitions determined in TDA follows closely, from below,  the one evaluated in RPA  and accounts almost entirely (from 90\% to 95\%) for  the RPA energy weighted sum (EWS) of all multipoles.  For the isovector $E1$, instead, it overestimates the RPA sum by $\sim 10\%$. 
This different behavior is consistent with the schematic model \cite{Rowe70}.

Also the STDA/EMPM and the SRPA integrated strengths evolve closely. Their smooth evolution  reflects the fragmentation of the strengths induced by the coupling to 2p-2h configurations. The STDA/EMPM running sum tends exactly to the asymptotic TDA EWS. Analogously, in SRPA and RPA, the EWS coincides in most nuclei and for most multipolarities. 

There are some exceptions. In $^{40}$Ca, the SRPA  underestimates appreciably the RPA $E3$ EWS. The simple reason is that the lowest level which was supposed to carry a large $E3$ strength is imaginary. Another peculiarity of SRPA, which was discussed in Ref.~\cite{Pap2014}, is that the RPA $E2$ EWS is preserved only if we include in the running sum the negative energy $2^+$ level and its strength.

A final remark concerns the effect of the center of mass. The EWS of all multipolarity remains unaltered whether we remove or not the c.m. motion. Due to this invariance, the conservation of the EWS can provide valuable guidance in any beyond mean-field extension.

 \section{Conclusion} 
  From the present survey we can draw some clearcut conclusions. 
 The differences between TDA and RPA  as well as between STDA  and SRPA responses are marginal, except for the octupole transitions where the RPA low-lying octupole peak is shifted considerably downward with respect to TDA. In going from RPA to SRPA, such a peak is pushed further down in energy and 
can disappear completely if the corresponding eigenvalue becomes imaginary.  This instability casts a shadow on the reliability of approaches which are meant to go beyond RPA.  SRPA then remains applicable primarily on higher-lying collective modes, {\em i.e.}, giant resonances. The TDA and STDA spectra do not exhibit any anomaly. 
  
In RPA, it is possible to remove almost completely the low-lying spurious isovector $E1$ peak thanks to the adoption of a HF basis combined with the subtraction of the c.m. coordinates from the dipole operator.  In TDA, the Gram-Schmidt orthogonalization procedure eliminates any spurious admixtures from both isoscalar and isovector responses. 

In going to  STDA and SRPA,  however, the c.m. motion affects all multipoles. Its spuriousness  spreads over the entire spectrum of each multipole thereby inducing non negligible  distortions  of the isoscalar $E1$, $E2$, and $E3$ strength functions.  Only the isovector $E1$ response is marginally affected.

 STDA and SRPA do not offer any recipe for  removing these distortions, while in the EMPM the joint use of Gram-Schmidt and SVD pins down and removes completely and exactly the spurious admixtures from all multipole responses.

 The EMPM is exactly identical to STDA within the space encompassing 1p-1h plus 2p-2h configuration under the simplifying assumption of neglecting the coupling between the HF and the 2p-2p basis states. However, the anomaly of the $^{48}$Ca spectrum, where the $2^+$ falls below the HF ground state, ratifies the failure of  STDA and SRPA in describing the spectroscopy of some nuclei, at least for the potential adopted here. The EMPM shows how to remove this anomaly. One should refer the levels to a fully correlated ground state by taking into account the HF to 2p-2h coupling jointly with 
 enlarging the configuration space so as to include the 3p-3h basis, as it was done in \cite{DEGREGORIO2021}. In other words, one should move to full shell model or to the EMPM.

The EMPM is more general than STDA and is exactly identical to shell model within a given configuration space. With respect to shell model, it is more involved
 but offers significant advantages. It allows for truncations of the multiphonon basis even if a large space including very  high energy configurations is adopted.  These are accounted for by the TDA phonons building up the $n$-phonon states.   It is suitable for investigating low energy spectra as in shell model but also the high energy responses as in RPA or SRPA. Last, but not least, the intrinsic motion is decoupled completely and exactly from the c.m. motion for any s.p. basis and under no restrictions.

   
\begin{acknowledgments}
This work was partly supported by the Czech Science Foundation (Czech Republic), P203-19-14048S and by the Charles University Research Center UNCE/SCI/013. The work of P.P. was supported by the Rare Isotope Science Project of the Institute for Basic Science funded by the Ministry of Science, ICT and Future Planning and the National Research Foundation (NRF) of Korea (2013M7A1A1075764). P.V. thank the INFN for financial support. Computational resources were provided by the CESNET LM2015042 and the CERIT Scientific Cloud LM2015085, under the program "Projects of Large Research, Development, and Innovations Infrastructures". 
 This work is co-funded by EU-FESR, PON Ricerca e Innovazione 2014-2020 - DM 1062/2021.
\end{acknowledgments}
\bibliographystyle{apsrev}
\bibliography{EMTDRPA}
\end{document}